\documentclass[twocolumn,amsmath,amssymb,aps,pra,10pt]{revtex4-1}
\usepackage{graphicx}% Include figure files
\usepackage{dcolumn}% Align table columns on decimal point
\usepackage{bm}% bold math
\newcommand{\bk}{\mathbf k}
\newcommand{\bq}{\mathbf q}
\newcommand{\mut}{\widetilde\mu}
\newcommand{\omsing}{\widetilde\omega}

\newcommand{\nF}{n^{}_{\rm F}}
\newcommand{\gs}{g^{}_{\rm s}}
\newcommand{\omp}{\omega^{}_{\rm p}}

\newcommand{\rA}{\mathrm A}
\newcommand{\rB}{\mathrm B}
\newcommand{\rC}{\mathrm C}
\newcommand{\rD}{\mathrm D}
\newcommand{\ut}{\widetilde u}
\newcommand{\qt}{\widetilde q}
\newcommand{\cF}{\mathcal F}

\DeclareMathOperator{\Tr}{Tr}
\DeclareMathOperator{\arctanh}{arctanh}
\DeclareMathOperator{\sgn}{sgn}
\DeclareMathOperator{\re}{Re}
\DeclareMathOperator{\im}{Im}

\usepackage[pdftex,bookmarks=false,hypertexnames=false,breaklinks=true,colorlinks=true,urlcolor=blue,citecolor=blue,linkcolor=blue]{hyperref}
\newcommand*{\doiref}[2]{\href{http://dx.doi.org/#1}{#2}}
\newcommand*{\arx}[1]{\href{http://arxiv.org/abs/#1}{arXiv: \!\!#1}}

\begin{document}

\title{Dynamical polarization and plasmons in a two-dimensional system with merging Dirac points}
\author{P. K. Pyatkovskiy and Tapash Chakraborty}
\affiliation{Department of Physics and Astronomy,
University of Manitoba, Winnipeg, Canada R3T 2N2}

\date{\today}
\begin{abstract}
We have studied the dynamical polarization and collective excitations in an anisotropic two-dimensional 
system undergoing a quantum phase transition with merging of two Dirac points. Analytical results for
the one-loop polarization function are obtained at the finite momentum, frequency, and chemical 
potential. The evolution of the plasmon dispersion across the phase transition is then analyzed within 
the random phase approximation. We derive analytically the long-wavelength dispersion of the
undamped anisotropic collective mode and find that it evolves smoothly at the critical merging
point. The effects of the van Hove singularity on the plasmon excitations are explored in detail.
\end{abstract}
%\pacs{73.21.-b, 71.45.Gm}

\maketitle

\section{Introduction}

For more than a decade, we have been witnessing the rise of a plethora of ever new two-dimensional (2D) 
materials displaying their unique electronic properties, which has initiated major activities in those 
systems. Leading the pack was, of course, monolayer and bilayer graphene displaying the behavior of 
``Dirac fermions'' of the charge carriers with ``Dirac points'' where the two energy bands
meet \cite{graphene_book,abergeletal} with linear dispersions in the vicinity that forms the characteristic
``Dirac cones'' \cite{nsreview}. Their many exotic physical properties, in particular, in a strong magnetic 
field, have been well documented, and range from the magnetic field effects in the extreme quantum
limit \cite{fqhe} to Hofstadter butterflies \cite{hofstadter,butterflies}. These were then followed by other 
graphenelike systems, such as silicene and germanene \cite{silicene,vadim,wenchen}, the 2D version of 
black phosphorus (BP) \cite{phosphorene,areg}, and, most recently, the planar electron systems in ZnO 
heterojunctions \cite{zno,wenchen_2}. Interestingly, an anisotropic two-dimensional system can undergo a 
transition between an insulating state with gapped spectrum and a semimetal state with two Dirac cones 
separated in the momentum space. The possibility of such a quantum phase transition has been considered 
theoretically in honeycomb lattice models \cite{hasegawa,zhu,wunsch,goerbig,pereira} and few-layer black phosphorus where the band inversion can be
induced by an external perpendicular electric field \cite{zunger,dolui,pereira_2} or by doping \cite{baik}. 
The gapless spectrum at the phase transition point may arise in the $\mathrm{TiO}_2/\mathrm{VO}_2$ 
nanostructures \cite{pardo,huang}. Experimentally, merging or creation of Dirac points has been observed 
in systems of ultracold atoms \cite{esslinger}, photonic crystals \cite{rechtsman}, microwave analog of 
graphene \cite{bellec}, and, more recently, in a potassium-doped few-layer BP \cite{kim}.

Various properties of a system undergoing this phase transition have been reported in the literature, 
which include the Landau levels and the Hofstadter spectrum \cite{dietl}, the Hall conductivity \cite{yuan2},
effects of disorder \cite{carpentier}, the quantum critical behavior \cite{isobe,cho}, and the transport
characteristics \cite{adroguer}. In this paper, we consider the dynamical polarization and collective excitations 
utilizing the model introduced in Ref.~\cite{goerbig} in which the phase transition is governed 
by a single parameter $\Delta$ that changes its sign across the critical point (Fig.~\ref{fig:merging}). 
This model provides a universal description \cite{goerbig} for a two-dimensional system in the vicinity of the phase 
transition with two merging Dirac points related by time-reversal symmetry.
Previous results related to our present study include the 
long-wavelength plasmon dispersion at the critical point ($\Delta=0$) obtained in Ref.~\cite{banerjee}, 
spectrum of collective excitations in a single- and few-layer BP \cite{low,rodin} (only the conduction 
band or the valence band was taken into account due to the large value of the gap), and the spectrum of 
plasmons across the phase transition obtained numerically within a tight-binding model for bilayer
BP \cite{jin}.

We calculate the one-loop dynamical polarization function at zero temperature for arbitrary values of 
the Fermi energy and the gap. In general, we are able to perform one momentum integration and derive 
an expression in terms of a single integral valid for arbitrary complex frequencies. This expression 
is used to study numerically the evolution of the plasmon dispersion across the phase transition within
the random phase approximation (RPA). The imaginary part of the vacuum polarization function and the 
long-wavelength spectrum of collective excitations are evaluated analytically.

\section{Polarization function}
\label{section:polarization}

\begin{figure}
\includegraphics[width=0.95\columnwidth]{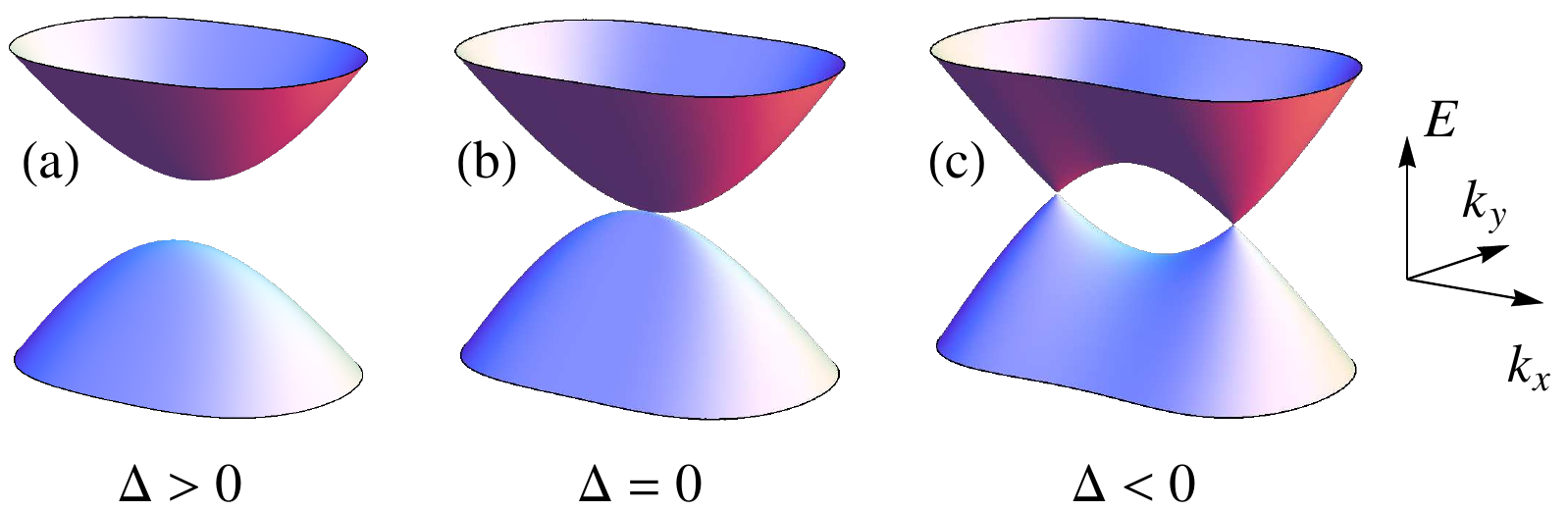}
\caption{Evolution of the electron energy spectrum at the phase transition.}
\label{fig:merging}
\end{figure}

We use a universal low-energy two-band Hamiltonian \cite{goerbig} describing the merging transition,
\begin{equation}
H=(\Delta+ak_x^2)\sigma^{}_x+vk^{}_y\sigma^{}_y,
\label{H}
\end{equation}
where the Pauli matrices $\sigma^{}_x$, $\sigma^{}_y$ act on the two-component wave functions. The 
spin-orbit coupling is neglected and the presence of two spin states is accounted for by the 
degeneracy factor $\gs=2$. The energy eigenvalues are given by
\begin{equation}
E_\bk^\lambda=\lambda\sqrt{(ak_x^2+\Delta)^2+v^2k_y^2}, \qquad \lambda=\pm.
\label{eig_energy}
\end{equation}
The Hamiltonian~(\ref{H}) can also be used to describe the single-layer BP
($\Delta\approx0.8$~eV) \cite{rudenko} when the difference in the effective masses
of the positive- and negative-energy bands is 
neglected. In the case when the chemical potential $\mu$ lies within the conductance band and the gap 
is large ($0<\mu-\Delta\ll\Delta$), we can neglect the contribution from the negative-energy band 
for energies close to the Fermi level and, at small momenta, approximate Eq.~(\ref{eig_energy}) by
\begin{equation}
E^+_\bk-\mu\approx\frac{k_x^2}{2m_x}+\frac{k_y^2}{2m^{}_y}-\mu^{}_0
\label{large_gap_spectrum}
\end{equation}
with $m^{}_x=1/(2a)$ and $m^{}_y=\Delta/v^2$ being the effective masses in the $x$ and $y$ directions, 
respectively, and the chemical potential $\mu^{}_0=\mu-\Delta$ measured from the bottom of the positive-energy band.
At the critical point, $\Delta=0$ [Fig.~\ref{fig:merging}(b)], the spectrum is linear in 
the $y$ direction, while quadratic in the $x$ direction (with the same effective mass $m^{}_x$). Such 
a system is often referred to in the literature as the ``semi-Dirac'' system.

In the case of $\Delta<0$, the spectrum has two Dirac cones [Fig.~\ref{fig:merging}(c)] located at
$\bk=(\pm K^{}_x,0)$ with $K^{}_x=\sqrt{-\Delta/a}$. In the vicinity of these points, the linearized 
Hamiltonian~(\ref{H}) reads
\begin{equation}
H\simeq \pm v^{}_x(k^{}_x\mp K^{}_x)\sigma^{}_x+vk^{}_y\sigma^{}_y,
\label{lin_H}
\end{equation}
where the velocities $v^{}_x=2\sqrt{-a\Delta}$ in the $x$ direction and $v$ in the $y$ direction are different in general. 
The spectrum has saddle points $E_{\bk=0}^\pm=\pm|\Delta|$ with divergent density of states 
(the van Hove singularity).

The one-loop polarization function at finite temperature $T$ is given by
\begin{align}
\Pi(i\omega^{}_m,\bq)&=\gs T\sum_{n=-\infty}^\infty\int\frac{d^2k}{(2\pi)^2}
\Tr\bigl[G(i\Omega^{}_n,\bk) \nonumber \\
&\quad\times G(i\Omega^{}_n+i\omega^{}_m,\bk+\bq)\bigr],
\end{align}
where $\omega^{}_m=2\pi mT$, $\Omega^{}_n=(2n+1)\pi T$ are the Matsubara frequencies, and the Green's 
function is 
\begin{equation}
G(i\Omega^{}_n,\bk)=\frac{i\Omega^{}_n+\mu+(ak_x^2+\Delta)\sigma^{}_x+vk^{}_y\sigma^{}_y}{(i\Omega^{}_n
+\mu)^2-(ak_x^2+\Delta)^2-v^2k_y^2}.
\end{equation}
Evaluating the trace and the sum over $n$ yields
\begin{equation}
\Pi(i\omega^{}_m,\bq)=\gs\!\int\!\!\frac{d^2k}{(2\pi)^2}\!\!\sum_{\lambda,\lambda'=\pm}\!\!\!F^{\lambda,
\lambda'}_{\bk,\bk+\bq} \frac{\nF(E_\bk^\lambda)-\nF(E_{\bk+\bq}^{\lambda'})}{E_\bk^\lambda-E_{\bk+
\bq}^{\lambda'}+i\omega^{}_m},
\label{Lindhard}
\end{equation}
where $\nF(x)=[e^{(x-\mu)/T}+1]^{-1}$ and
\begin{equation}
F^{\lambda,\lambda'}_{\bk,\bk'}=\frac12\biggl[1
+\frac{(\Delta+ak_x^2)(\Delta+ak_x'^2)+v^2k^{}_yk_y'}{E_\bk^\lambda E_{\bk'}^{\lambda'}}\biggr]
\end{equation}
is the wave-function overlap factor. In the following, we consider only the case of $T=0$, when
$\nF(x)\to\theta(\mu-x)$ and the polarization function can be written as the sum of two terms,
\begin{equation}
\Pi(i\omega,\bq)=\Pi^{}_0(i\omega,\bq)+\Pi^{}_1(i\omega,\bq),
\end{equation}
where $\Pi^{}_0(i\omega,\bq)$ is the ``vacuum'' polarization at $\mu=0$ and $\Pi^{}_1(i\omega,\bq)$ gives an
additional contribution when $\mu>\Delta$ (we choose $\mu\geqslant0$, and the case $\mu<0$ is equivalent
because of the electron-hole symmetry). These two terms are given by
\begin{align}
\Pi^{}_0(i\omega,\bq)&=-\chi_\infty^-(i\omega,\bq), \nonumber \\
\Pi^{}_1(i\omega,\bq)&=\chi_\mu^+(i\omega,\bq)+\chi_\mu^-(i\omega,\bq),
\end{align}
where
\begin{equation}
\chi_\mu^\pm(\omega,\bq)=\gs\int\frac{d^2k}{(2\pi)^2}\sum^{}_{\sigma=\pm}
\frac{\theta(\mu-E^+_\bk)F^{+,\pm}_{\bk,\bk+\bq}}{E^+_\bk-E^\pm_{\bk+\bq}+\sigma\omega}.
\label{chi}
\end{equation}
The $k^{}_y$ integration in the above equation can be performed analytically for an arbitrary complex 
frequency $\omega$ away from the real axis (see Appendix~\ref{appendix:calc}). The resulting expressions are
\begin{widetext}
\begin{equation}
\begin{split}
\Pi^{}_0(\omega,\bq)&=-\frac{\gs}{2\pi^2v(v^2q_y^2-\omega^2)^2}\int\limits_{-\infty}^\infty
dk^{}_x\Biggl[v^2q_y^2(v^2q_y^2-\omega^2)
+\eta\xi(v^2q_y^2+\omega^2)\re\bigl[\arctanh(\xi/\eta)\bigr] \\
&\quad-\biggl(\xi^2\omega^2\frac{\sqrt{\alpha-\omega^2}}{\sqrt{\beta-\omega^2}}
+v^2q_y^2\eta^2\frac{\sqrt{\beta-\omega^2}}{\sqrt{\alpha-\omega^2}}\biggr)
\arctanh\frac{\sqrt{\min(\alpha,\beta)-\omega^2}}{\sqrt{\max(\alpha,\beta)-\omega^2}}\Biggr],
\end{split}
\label{Pi0}
\end{equation}
\begin{equation}
\begin{split}
\Pi^{}_1(\omega,\bq)&=\frac{\gs\theta(\mu-\Delta)}{4\pi^2v(v^2q_y^2-\omega^2)^2}\int\limits_{-\sqrt{(\mu-\Delta)/a}}^{\sqrt
{(\mu-\Delta)/a}}dk^{}_x
\theta(\mu+\Delta+ak_x^2)\biggl[\eta\xi(v^2q_y^2+\omega^2)\arctanh\frac{\mut}{\mu} \\
&\quad-\biggl(\xi^2\omega^2\frac{\sqrt{\alpha-\omega^2}}{\sqrt{\beta-\omega^2}}
+v^2q_y^2\eta^2\frac{\sqrt{\beta-\omega^2}}{\sqrt{\alpha-\omega^2}}\biggr)
\arctanh\frac{\mut(\eta\xi+v^2q_y^2-\omega^2)-2vq^{}_y(ak_x^2+\Delta)^2}
{\mu\sqrt{\alpha-\omega^2}\sqrt{\beta-\omega^2}} \\
&\quad-2vq^{}_y\eta\xi\omega\arctanh\frac{\eta\xi+v^2q_y^2-\omega^2+2vq^{}_y\mut}{2\mu\omega}+(q^{}_y
\to-q^{}_y)\biggr],
\end{split}
\label{Pi1}
\end{equation}
\end{widetext}
where
\begin{align}
\xi&=aq^{}_x(q^{}_x+2k^{}_x), \nonumber \\
\eta&=\xi+2(ak_x^2+\Delta), \nonumber \\
\alpha&=\eta^2+v^2q_y^2, \label{eta_xi_alpha_beta_def} \\
\beta&=\xi^2+v^2q_y^2, \nonumber \\
\mut&=\sqrt{\mu^2-(ak_x^2+\Delta)^2}, \nonumber
\end{align}
and the retarded polarization on the real $\omega$ axis is obtained using the prescription 
$\omega\to\omega+i0$. We use Eqs.~(\ref{Pi0}) and~(\ref{Pi1}) in our numerical calculations of the collective 
excitation spectrum and also to analytically obtain some important limits.

The imaginary part of the vacuum term can be calculated analytically (see Appendix~\ref{appendix:imPi0}).
If $\Delta<0$ and $q^{}_x<2K^{}_x$, it has a logarithmic singularity at $\omega=\pm\omsing$, where
\begin{equation}
\omsing=\sqrt{v^2q_y^2+a^2(2K_x^2-q_x^2/2)^2}.
\end{equation}
In the vicinity of this singularity, $\im\Pi^{}_0(\omega,\bq)$ is given by (including terms finite 
at $\omega=\pm\omsing$)
\begin{align}
&\im\Pi^{}_0(\omega,q^{}_x,q^{}_y)\simeq\frac{\mp\gs}{64\pi\sqrt av}\Biggl[\frac{v^2q_y^2}{a^{3/2}\qt^3}
\ln\frac{512a^2\qt^6}{\omsing|q_x^2-\qt^2||\omega\mp\omsing|} \nonumber \\
&\quad-\frac{\omsing^2(q_x^2-3\qt^2)+v^2q_y^2(q_x^2+\qt^2)}{a^{3/2}q_x^2\qt^3} \\
&\quad+\frac{\omsing^2-8a^2(q_x^4+2q_x^2\qt^2-\qt^4)}{a^{3/2}q_x^3}\ln\frac{|q^{}_x-\qt|}{q^{}_x+\qt}
\Biggr], \nonumber
\end{align}
where $\qt=\sqrt{K_x^2/2-q_x^2/8}$. For $q^{}_x=0$, this simplifies to
\begin{equation}
\im\Pi^{}_0(\omega,0,q^{}_y)\simeq\frac{\mp\gs vq_y^2}{\pi\sqrt a(-8\Delta)^{3/2}}
\biggl(\ln\frac{128\Delta^2}{\omsing^{}_0|\omega\mp\omsing^{}_0|}-\frac83\biggr)
\end{equation}
with $\omsing^{}_0=\sqrt{4\Delta^2+v^2q_y^2}$. 
This logarithmic divergence for $\Delta<0$ is due to the van Hove singularity which results in the saddle point in the
interband single-particle excitation (SPE) energy $E_\bk^+-E_{\bk+\bq}^-$ [the real frequency 
corresponding to the pole of the integrand in Eq.~(\ref{Lindhard})] as a function of $\bk$ for a given
external wave vector $\bq$. A similar divergence of $\im\Pi^{}_0(\omega,\bq)$ due to the
presence of the van Hove singularity appears in graphene \cite{stauber,yuan}. In contrast to
the case of graphene, in our model this singularity occurs only at a single point in the momentum space between the two Dirac cones. Because of this, the
divergent term is proportional to $q^{}_y$ and vanishes for the momentum directed along the $x$ axis. 

For $\Delta=0$, some limiting cases of vacuum polarization~(\ref{Pi0}) can be evaluated analytically,
\begin{equation}
\Pi(\omega,q_x=0,q_y)=-\frac{\Gamma(5/4)}{\Gamma(3/4)}\frac{\gs vq_y^2}{6\sqrt{\pi a}(v^2q_y^2-\omega^2)^{3/4}},
\end{equation}
\begin{equation}
\Pi(\omega=0,q_x,q_y=0)=-\frac{\gs|q_x|}{16v},
\end{equation}
in agreement with the previously reported results \cite{isobe,cho}.

\section{Plasmons}
\label{section:plasmons}

Plasmon dispersion $\omp(\bq)$ in the RPA is obtained from zeros of the dielectric function
\begin{equation}
\epsilon(\omega,\bq)=1-V(\bq)\Pi(\omega,\bq),
\label{dielectric_function}
\end{equation}
where $\Pi(\omega,\bq)$ is the one-loop polarization function and $V(\bq)=2\pi e^2/(\kappa q)$ 
is the Coulomb potential screened only by the substrate with the corresponding background dielectric 
constant $\kappa$.

We numerically found the real solutions $\omega=\omp(\bq)$ of Eq.~(\ref{dielectric_function}) in
the regions where $\im\Pi(\omega,\bq)=0$, i.e., the Landau damping is absent. For the
SPE regions where the imaginary part of the polarization function is nonzero, we calculate the energy-loss
function $-\im[1/\epsilon(\omega,\bq)]$, the peaks of which represent the damped plasmons. Our
approach assumes the strictly two-dimensional system and does not take into account the charge
distribution in the perpendicular direction. Nevertheless, our results will be valid if the
characteristic length of this distribution $l^{}_z$ (e.g., the interlayer distance in the case of 
bilayer BP) is much smaller then $1/|\bq|$ for the considered wave vectors $\bq$.  

\begin{figure}
\includegraphics[width=0.92\columnwidth]{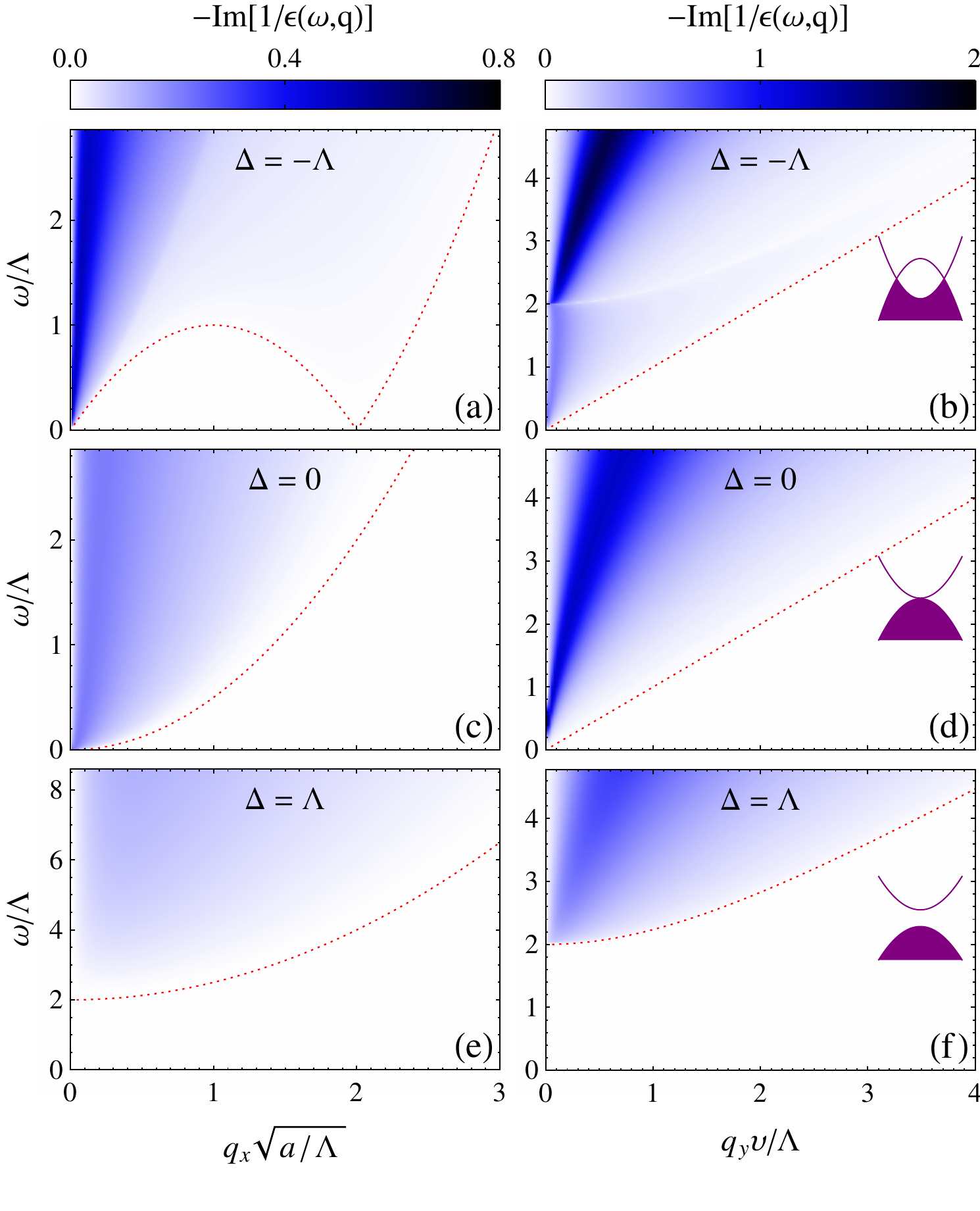}
\caption{Energy-loss function $-\im[1/\epsilon(\omega,\bq)]$ at $\mu=0$ for positive, zero, and
negative $\Delta$. Left: $q^{}_y=0$, $\kappa v=10^{-3}c$, and $\Lambda$ is an arbitrary energy
scale. Right: $q^{}_x=0$, $2\kappa\sqrt{a\mu}=10^{-3}c$, and $\kappa^2\Lambda=4\times10^{-3}e^2/a$. 
The boundaries of the SPE regions are marked by dotted lines. The insets in the right panels schematically show the band structure and filling of the bands.}
\label{fig:pl_vac}
\end{figure}

In the case of $\mu=0$, there are no real solutions of Eq.~(\ref{dielectric_function}) and the
energy-loss function is shown in Fig.~\ref{fig:pl_vac}. For $\Delta<0$, the logarithmic divergence due 
to the van Hove singularity manifests itself as a dip in $-\im[1/\epsilon(\omega,\bq)]$ for $q^{}_y\ne0$ 
and $\omega=\omsing$ followed by a peak at a larger energy [Fig.~\ref{fig:pl_vac}(b)]. Analogous behavior 
has also been reported in graphene \cite{yuan}.

\begin{figure}[!ht]
\includegraphics[width=\columnwidth]{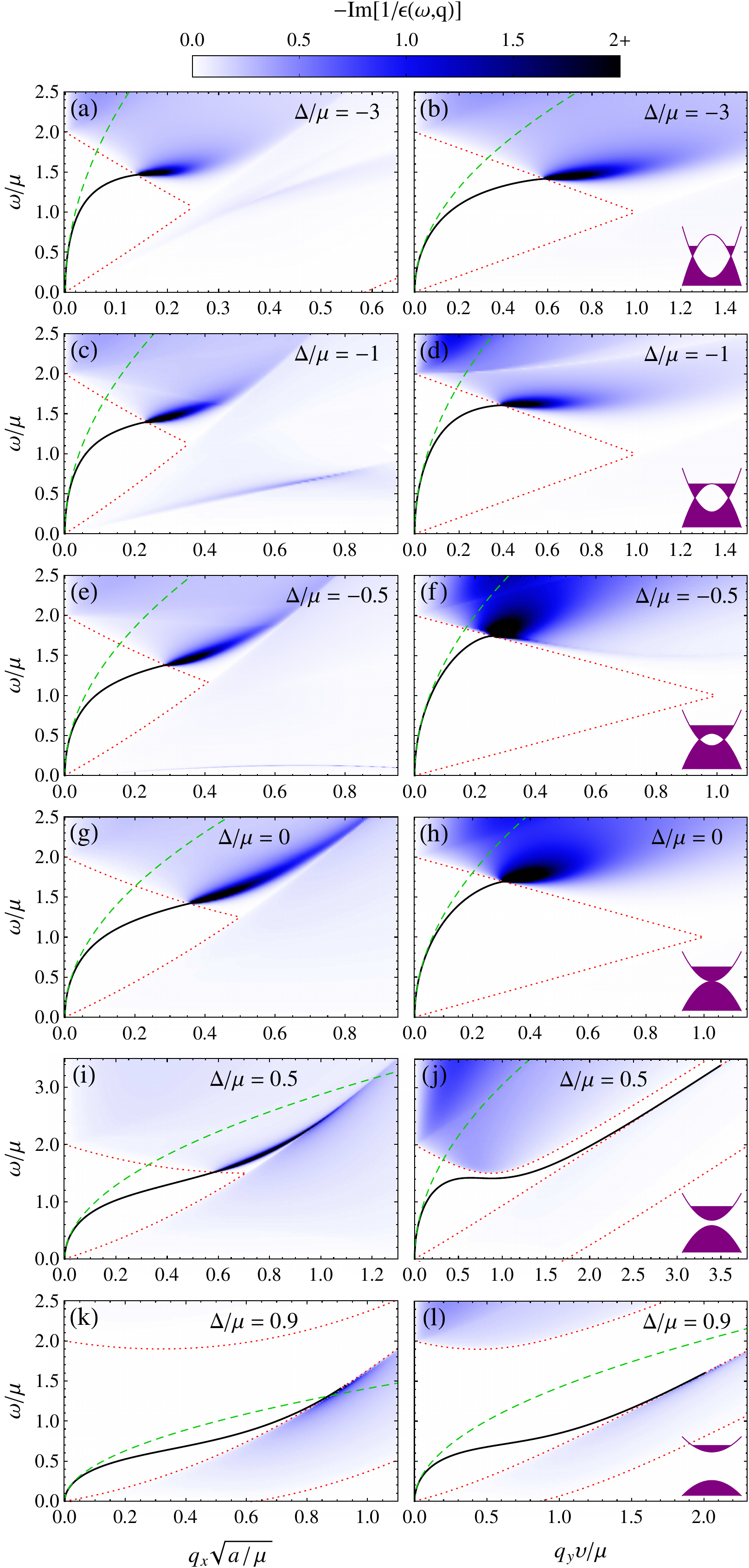}
\caption{Energy-loss function $-\im[1/\epsilon(\omega,\bq)]$ at $q^{}_y=0$, $\kappa v=10^{-3}c$
(left) and $q^{}_x=0$, $2\kappa\sqrt{a\mu}=10^{-3}c$ (right) for different values of $\Delta/\mu$. 
The undamped plasmon mode and its long-wavelength asymptote~(\ref{plasm_disp_small_q}) are shown 
by solid and dashed lines, respectively. The boundaries of the SPE regions are marked by dotted lines. The insets in the right panels schematically show the band structure and filling of the bands.}
\label{fig:pl_mu}
\end{figure}

The evolution of the plasmon spectrum across the phase transition at nonzero chemical potential
is shown in Fig.~\ref{fig:pl_mu}. The momentum is chosen to be aligned with one of the
principal axes. In the case of $q^{}_y=0$, the dielectric function~(\ref{dielectric_function})
expressed in terms of the dimensionless momentum $q^{}_x\sqrt{a/\mu}$ and energy $\omega/\mu$
depends only on a single adjustable parameter $\kappa v$. Similarly, $\epsilon(\omega,\bq)$ at
$q^{}_x=0$ can be represented as a function of the dimensionless momentum $q^{}_yv/\mu$ and energy
$\omega/\mu$, which depends on a single parameter $\kappa^2a\mu$. In our numerical calculations,
we choose the values of the parameters $\kappa v=10^{-3}c\approx3\times10^{5}$~m/s, $a\mu=v^2/4$.
The latter choice corresponds, e.g., to $m^{}_x=1/(2a)=m^{}_{\rm e}$ and $\mu\approx0.26$~eV, where
$m^{}_{\rm e}$ is the bare electron mass.

\begin{figure}
\includegraphics[width=0.65\columnwidth]{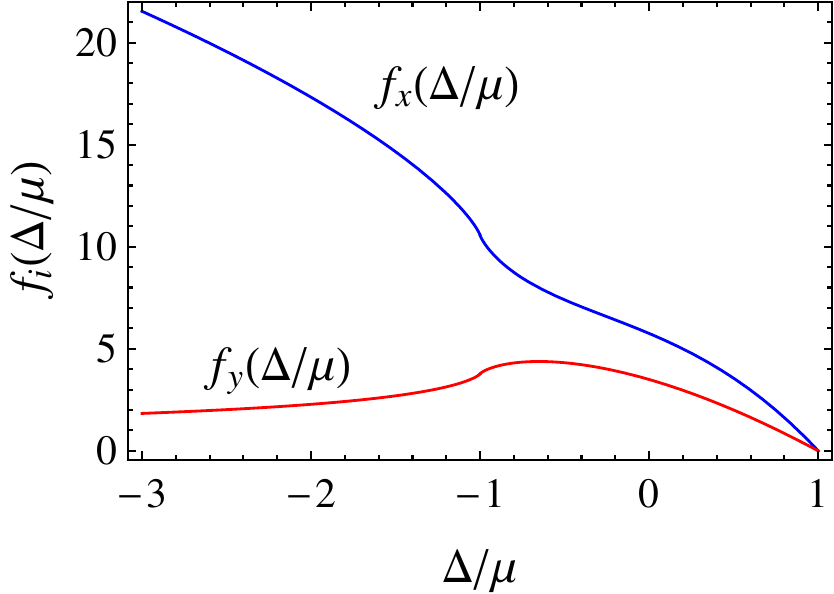}
\caption{Factors $f^{}_x(\Delta/\mu)$ and $f^{}_y(\Delta/\mu)$ determining the long-wavelength 
plasmon dispersion~(\ref{plasm_disp_small_q}).}
\label{fig:fxfy}
\end{figure}

In the regime $\sqrt aq^{}_x,vq^{}_y\ll\omega\ll\mu$, i.e., relevant for the long-wavelength plasmons, 
the asymptotic behavior of the polarization function is
\begin{equation}
\Pi(\omega,\bq)=\frac{\gs\sqrt\mu}{4\pi^2v\sqrt a\omega^2}\bigl[\mu
aq_x^2f^{}_x(\Delta/\mu)+v^2q_y^2f^{}_y(\Delta/\mu)\bigr],
\label{pol_long_wavelength}
\end{equation}
where the functions $f^{}_{x,y}(\delta)$ are defined as
\begin{equation}
\begin{split}
f^{}_x(\delta)&=8\int_{t^{}_0}^1dt\frac{t^2\sqrt{t-\delta}}{\sqrt{1-t^2}}, \\
f^{}_y(\delta)&=2\int_{t^{}_0}^1dt\frac{\sqrt{1-t^2}}{\sqrt{t-\delta}},
\end{split}
\label{fxfy}
\end{equation}
with $t^{}_0=\max(\delta,-1)$ and shown in Fig.~\ref{fig:fxfy}. The analytical expressions for
$f^{}_{x,y}(\delta)$ in terms of the complete elliptic integrals are given in Appendix~\ref{appendix:fxfy}.

Using Eq.~(\ref{pol_long_wavelength}), we obtain the long-wavelength plasmon dispersion,
\begin{align}
\omp(\bq)&\simeq\sqrt{\frac{\gs e^2\mu q}{2\pi\kappa}}
\biggl[\frac{\sqrt{\mu a}}vf^{}_x(\Delta/\mu)\cos^2\theta \nonumber \\
&\quad+\frac v{\sqrt{\mu a}}f^{}_y(\Delta/\mu)\sin^2\theta\biggr]^{1/2},
\label{plasm_disp_small_q}
\end{align}
where $\cos\theta=q^{}_x/q$, $\sin\theta=q^{}_y/q$. It has the usual square-root dependence on the 
momentum just as expected for a 2D system, with the anisotropy of the spectrum being fully determined by the 
dimensionless factor $\sqrt{\mu a}/v$ and the two functions $f^{}_{x,y}(\Delta/\mu)$. This undamped 
plasmon mode lies in the gap between the interband and the intraband SPE regions~(Fig.~\ref{fig:pl_mu}).

For $-\Delta\gg\mu$, i.e., for well separated Dirac cones [Figs.~\ref{fig:pl_mu}(a) and~\ref{fig:pl_mu}(b)], using the 
asymptotics
\begin{equation}
\begin{split}
f^{}_x(\delta)&=4\pi|\delta|^{1/2}+\mathcal O(|\delta|^{-1/2}), \\
f^{}_y(\delta)&=\pi|\delta|^{-1/2}+\mathcal O(|\delta|^{-3/2}), \qquad
-\delta\gg1,
\end{split}
\end{equation}
one can obtain the plasmon dispersion from Eq.~(\ref{plasm_disp_small_q}) in this limit
\begin{equation}
\omp(\bq)\simeq\sqrt{\frac{\gs e^2\mu q}{\kappa}}
\biggl[\frac{v^{}_x}{v}\cos^2\theta+\frac{v}{v^{}_x}\sin^2\theta\biggr]^{1/2}.
\end{equation}
This result corresponds to the linearized Hamiltonian~(\ref{lin_H}) and provides a generalization of 
the long-wavelength plasmon spectrum in a single-layer graphene \cite{shung} to the case of the different 
Fermi velocities in the $x$ and $y$ direction.

When the Fermi level crosses the van Hove singularity at $|\mu|=-\Delta$ and the Fermi surfaces of the two Dirac cones merge into a single one [the plasmon spectrum at this transition is shown in Figs.~\ref{fig:pl_mu}(c) and~\ref{fig:pl_mu}(d)], 
the plasmon frequency in the long wavelengths changes continuously as a function of $\Delta/\mu$ but 
has a logarithmic singularity of its derivative, as seen in Fig.~\ref{fig:fxfy}. The functions 
~(\ref{fxfy}) in the vicinity of this crossing are
\begin{equation}
\begin{split}
f^{}_x(-1+\varepsilon)&=2\sqrt2\bigl(\tfrac{56}{15}+\varepsilon\ln|\varepsilon|\bigr)+
\mathcal O(\varepsilon), \\
f^{}_y(-1+\varepsilon)&=\sqrt2\bigl(\tfrac83-\varepsilon\ln|\varepsilon|\bigr)+\mathcal 
O(\varepsilon).
\end{split}
\end{equation}
For $\Delta<0$, there also exists an additional damped plasmon mode in the $x$ direction for momenta
$0<q_x<2K_x$ with its maximum at $q^{}_x\sim K^{}_x$, which lies entirely in the intraband SPE region
[Figs.~\ref{fig:pl_mu}(a), \ref{fig:pl_mu}(c), and~\ref{fig:pl_mu}(e)].

At the crossing of the critical point $\Delta=0$ [Figs.~\ref{fig:pl_mu}(g) and~\ref{fig:pl_mu}(h)] the spectrum of 
the undamped plasmon changes smoothly and we have
\begin{equation}
\begin{split}
f^{}_x(0)&=3\sqrt\pi\,\Gamma(3/4)/\Gamma(9/4)\approx5.751, \\
f^{}_y(0)&=2\sqrt\pi\,\Gamma(5/4)/\Gamma(7/4)\approx3.496,
\end{split}
\end{equation}
in agreement \cite{note} with Ref.~\cite{banerjee}.

The effect of a small positive $\Delta/\mu$ is similar to that in the case of gapped graphene \cite{wang}: 
the plasmon mode becomes extended to larger values of momenta due to the opening of the gap between 
the interband and intraband SPE continua [Figs.~\ref{fig:pl_mu}(i) and~\ref{fig:pl_mu}(j)]. In the regime $\Delta\to\mu$ 
[Figs.~\ref{fig:pl_mu}(k) and~\ref{fig:pl_mu}(l)], electrons have approximately parabolic anisotropic dispersion
~(\ref{large_gap_spectrum}). Using the asymptotic behavior of~(\ref{fxfy}) for $\delta\to1$,
\begin{equation}
\begin{split}
f^{}_x(1-\varepsilon)&=2\sqrt2\pi\varepsilon+\mathcal O(\varepsilon^2), \\
f^{}_y(1-\varepsilon)&=\sqrt2\pi\varepsilon+\mathcal O(\varepsilon^2), \qquad 0\leqslant\varepsilon\ll1,
\end{split}
\end{equation}
we recover from Eq.~(\ref{plasm_disp_small_q}) in this limit,
\begin{equation}
\omp(\bq)\simeq\sqrt{\frac{\gs e^2\mu^{}_0q}{\kappa}}
\biggl[\sqrt{\frac{m^{}_y}{m^{}_x}}\cos^2\theta
+\sqrt{\frac{m^{}_x}{m^{}_y}}\sin^2\theta\biggr]^{1/2},
\end{equation}
as previously reported in Ref.~\cite{rodin} for the monolayer BP.

\section{Conclusions}
\label{section:Conclusions}

We have evaluated the polarization function and the spectrum of collective excitations in the 
two-dimensional system undergoing a topological phase transition with two merging Dirac points. 
A single integral representation for $\Pi(\omega,\bq)$ has been derived which is suitable for 
calculations on both real and imaginary frequency axes. An analytic expression was obtained 
for the imaginary part of the vacuum polarization and its asymptotic behavior near the logarithmic 
divergence due to the van Hove singularity. We analytically found the long-wavelength plasmon dispersion 
and numerically studied the spectrum of collective excitations for arbitrary momenta for both 
zero and nonzero values of the chemical potential. By evaluating the energy-loss function, we have
found both undamped and damped plasmon excitations at zero temperature and studied their evolution
across the merging transition. The presence of the van Hove singularity in the electron spectrum
leads to the existence of the gapped damped plasmon mode at zero chemical potential in the
semimetal phase. At finite $\mu$, there is one undamped anisotropic collective mode with the
square-root dispersion, which lies in the gap between the interband and intraband SPE regions. In
the gapped phase ($\Delta>0$), this undamped mode is generically extended to larger values of
momenta due to the enhanced separation between the two SPE continua. At the critical point
($\Delta=0$), the undamped plasmon dispersion changes smoothly, while an additional damped and
strongly anisotropic mode emerges at $\Delta<0$ in the interband SPE continuum. The crossing of
van Hove singularity by the Fermi level manifests itself in a divergent derivative of the
long-wavelength plasmon frequency.

\section*{Acknowledgments}
The work has been supported by the Canada Research Chairs Program of the Government of Canada.

\appendix

\section{Calculation of the polarization function}
\label{appendix:calc}

The integrand in Eq.~(\ref{chi}) can be written as
\begin{equation}
\sum^{}_{\sigma=\pm}
\frac{F^{+,\pm}_{\bk,\bk+\bq}}{E^+_\bk-E^\pm_{\bk+\bq}+\sigma\omega}
=\frac\partial{\partial k^{}_y}\cF^{}_\pm(\omega,\bq,\bk),
\label{derivative}
\end{equation}
where
\begin{align}
&\cF^{}_\pm(\omega,\bq,\bk)=\frac1{2v(v^2q_y^2-\omega^2)^2}\biggl[
-v^2q^{}_y\gamma(v^2q_y^2-\omega^2) \nonumber \\
&\quad+\eta\xi(v^2q_y^2+\omega^2)\arctanh\frac{2k^{}_y-q^{}_y}\gamma \nonumber \\
&\quad-2vq^{}_y\eta\xi\omega\arctanh\frac{\eta\xi-\omega^2+2v^2q^{}_yk^{}_y}{v\gamma\omega} \nonumber \\
&\quad-\biggl(\xi^2\omega^2\frac{\sqrt{\alpha-\omega^2}}{\sqrt{\beta-\omega^2}}
+v^2q_y^2\eta^2\frac{\sqrt{\beta-\omega^2}}{\sqrt{\alpha-\omega^2}}\biggr) \nonumber \\
&\quad\times\arctanh\frac{(2k^{}_y-q^{}_y)(\eta\xi+v^2q_y^2-\omega^2)-q^{}_y(\eta-\xi)^2}
{\gamma\sqrt{\alpha-\omega^2}\sqrt{\beta-\omega^2}} \nonumber \\
&\quad\mp\bigl(q^{}_y\to-q^{}_y,\;\xi\to-\xi\bigr)\biggr],
\end{align}
variables $\eta$, $\xi$, $\alpha$, $\beta$ are defined in Eq.~(\ref{eta_xi_alpha_beta_def}) and
\begin{equation}
\gamma=\sqrt{(\eta-\xi)^2/v^2+(2k^{}_y-q^{}_y)^2}.
\end{equation}
The function $\cF^{}_\pm(\omega,\bq,\bk)$ does not have any singularities for $\im\omega\ne0$
and the multivalued functions taken on their principal branches. Therefore, the definite
$k^{}_y$ integral of~(\ref{derivative}) is obtained straightforwardly by evaluating
$\cF^{}_\pm(\omega,\bq,\bk)$ in the integration limits, which, after some algebra, yields 
Eqs.~(\ref{Pi0}) and~(\ref{Pi1}).

\begin{figure}
\includegraphics[width=\columnwidth]{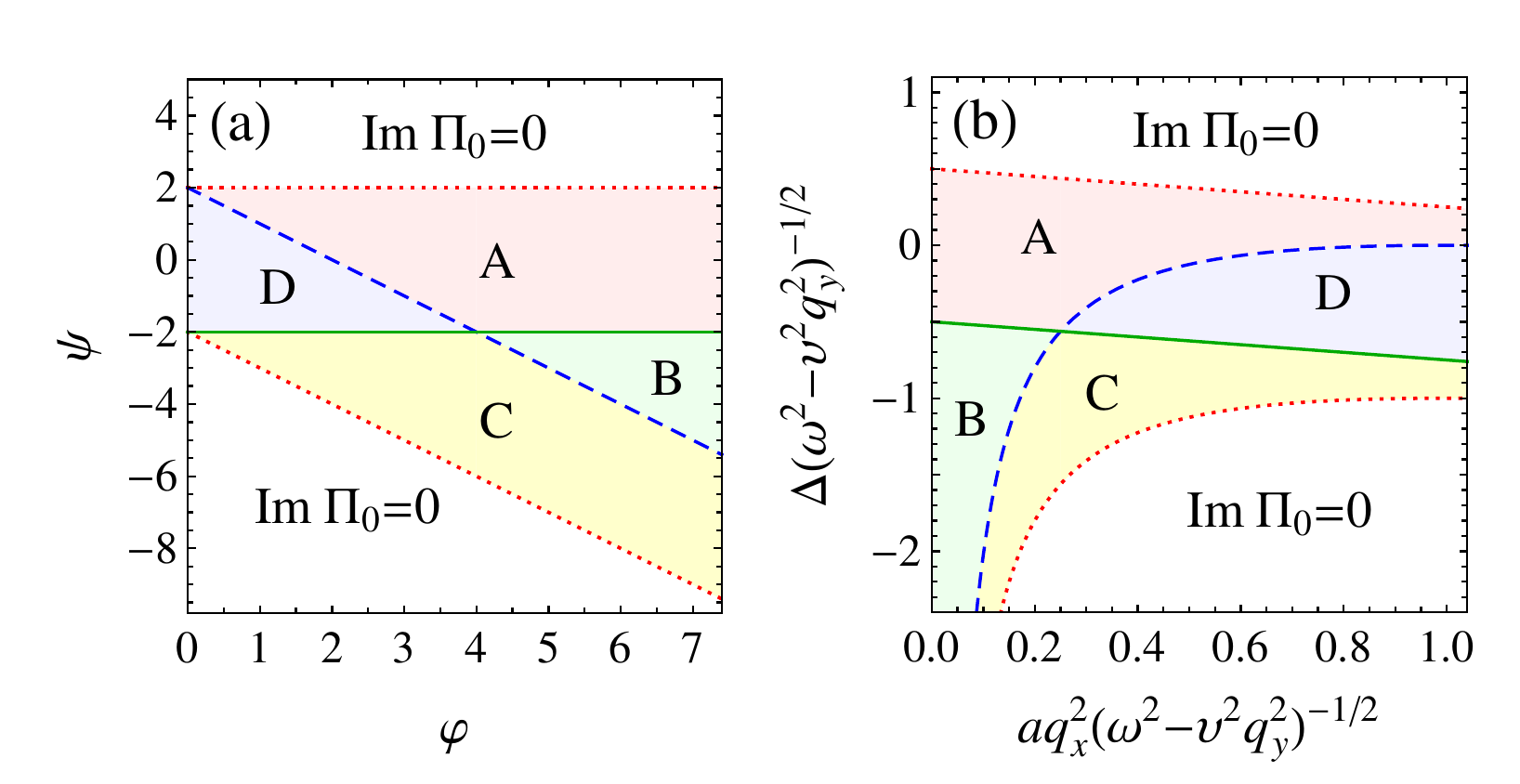}
\caption{Regions~(\ref{regions}) plotted for parameters (a) $\phi$ and $\psi$ or (b) $\bq$, $\omega$, 
and $\Delta$. At the boundaries, $\im\Pi_0(\omega,\bq)$ has a logarithmic singularity (solid line), 
logarithmically divergent derivative (dashed line) or jump discontinuity of the derivative (dotted line).}
\label{fig:regions}
\end{figure}

\section{Analytic expression for $\im\Pi^{}_0(\omega,\bq)$}
\label{appendix:imPi0}

The nonzero imaginary part of the expressions~(\ref{Pi0}) and~(\ref{Pi1}) for the polarization function 
originates from the regions where the argument of inverse hyperbolic tangent is real and larger than unity,
\begin{equation}
\im\bigl[\arctanh(x\pm i0)\bigr]=\pm\frac\pi2\theta(x^2-1).
\end{equation}
The step function above determines the integration limits which, for the nonvacuum term~(\ref{Pi1}),
involves the roots of the higher-order polynomials that cannot be written in a closed form in the general
case (for nonzero $q^{}_x$, $q^{}_y$, and $\Delta$). For the vacuum term, on the other hand, the imaginary 
part can be evaluated in terms of the complete elliptic integrals \cite{table}:
\begin{equation}
\begin{split}
&\im\Pi^{}_0(\omega,q^{}_x,q^{}_y)=-\frac{\gs\sqrt{|\omega|}\,\theta(1-\tau)\theta(2-\psi)
\theta(\nu)}{128\pi\sqrt a v(1-\tau)^{3/4}} \\
&\quad\times\sgn(\omega)\biggl\{4(2\psi+3\varphi-\tau\varphi+2\tau\psi)h^{}_iE(t^{}_i) \\
&\quad+(1-b)h_i^{-1}\Bigl[r^{}_iK(t^{}_i)+s^{}_i\bigl((16-4\psi\varphi)(1-\tau) \\
&\quad-\varphi^2(3-\tau)\bigr)\Pi(\pi/2,\rho^{}_i,t^{}_i)\Bigr]
\biggr\},
\end{split}
\label{imPi0}
\end{equation}
where
\begin{align}
d&=4\Delta(\omega^2-v^2q_y^2)^{-1/2}, &&\tau=v^2q_y^2/\omega^2, \nonumber \\
\varphi&=(\omega^2-v^2q_y^2)^{1/2}/(aq_x^2), &&\psi=d+1/\varphi, \\
\nu&=(\psi+\varphi+2)/4, &&b=(2-\psi)/\varphi, \nonumber
\end{align}
and the subscript $i=\rA,\rB,\rC,\rD$ determines the region in the ($\varphi,\psi$) space (see 
Fig.~\ref{fig:regions}):
\begin{equation}
\begin{split}
\rA:\qquad &-2<\psi<2, \quad \varphi>2-\psi, \\
\rB:\qquad &\psi<-2, \quad \varphi>2-\psi, \\
\rC:\qquad &\psi<-2, \quad -2-\psi<\varphi<2-\psi, \\
\rD:\qquad &-2<\psi<2, \quad 0<\varphi<2-\psi,
\end{split}
\label{regions}
\end{equation}
and
\begin{equation}
\begin{split}
&\rho^{}_\rA=\rho_\rD^{-1}=b, \qquad \rho^{}_\rC=\rho_\rB^{-1}=\nu,
\\
&t^{}_\rA=t^{}_\rC=t_\rB^{-1}=t_\rD^{-1}=\sqrt{b\nu},
\\
&s^{}_\rC=s^{}_\rD=1, \qquad s^{}_\rA=s^{}_\rB=-1,
\\
&h^{}_\rA=h^{}_\rC=1, \qquad h^{}_\rB=h^{}_\rD=\sqrt{b\nu},
\\
&r^{}_\rA=-\varphi\bigl[12+4\psi+3\varphi-\tau(4+4\psi+\varphi)\bigr],
\\
&r^{}_\rB=4\nu\bigl[2\psi-3\varphi+\tau(2\psi+\varphi)\bigr],
\\
&r^{}_\rC=-4\bigl[4+3\varphi-\tau(4+\varphi)\bigr],
\\
&r^{}_\rD=(\psi-2)(8+2\psi+3\varphi) \\
&\qquad+\tau[\psi(4+2\psi-\varphi)+2(8+\varphi)].
\end{split}
\end{equation}

\noindent
For $q^{}_x=0$, only regions A and B survive and Eq.~(\ref{imPi0}) simplifies to
\begin{equation}
\begin{split}
&\im\Pi^{}_0(\omega,0,q^{}_y)=-\frac{\gs\tau\sqrt{|\omega|}\theta(1-\tau)\theta(2-d)\sgn(\omega)}{12\pi\sqrt 
av(1-\tau)^{3/4}}  \\
&\;\times\!\left\{
\begin{array}{ll}
(1-d)K(\ut)+2dE(\ut), \quad & |d|<2, \medskip \\ 
\dfrac{2+d^2}{2\ut}K(1/\ut)+\dfrac{d(2-d)}{2\ut}E(1/\ut), \quad & d<-2,
\end{array}
\right.
\label{imPi0y}
\end{split}
\end{equation}
where $\ut=\sqrt{2-d}/2$.

\vfill

\onecolumngrid

\vspace{2cm}

\section{Expressions for $f^{}_{x,y}(\Delta/\mu)$}
\label{appendix:fxfy}

The functions~(\ref{fxfy}) determining the long-wavelength plasmon frequency can be written in 
terms of the complete elliptic integrals as
\begin{align}
f^{}_x(\delta)&=\frac{8\sqrt2}{15}\left\{
\begin{array}{ll}
2(9-2\delta^2)E(u)-(1+\delta)(9-2\delta)K(u), \qquad & |\delta|<1, \medskip \\ 
2u\bigl[(9-2\delta^2)E(1/u)+2\delta(1+\delta)K(1/u)\bigr], \qquad & \delta<-1,
\end{array}
\right. \\ & \nonumber \\
f^{}_y(\delta)&=\frac{4\sqrt2}{3}\left\{
\begin{array}{ll}
(1+\delta)K(u)-2\delta E(u), \qquad & |\delta|<1, \medskip \\ 
2u\bigl[(1+\delta)K(1/u)-\delta E(1/u)\bigr], \qquad & \delta<-1,
\end{array}
\right.
\end{align}
where $u=\sqrt{(1-\delta)/2}$. According to the notations used in Ref.~\cite{adroguer}, 
\begin{equation}
f^{}_x(\delta)=4\,\mathcal I^{}_3(0,\delta), \qquad
f^{}_y(\delta)=\mathcal I^{}_2(0,\delta).
\end{equation}
\twocolumngrid

\end{document}